\documentclass[english]{article}
\usepackage[latin9]{inputenc}
\usepackage{geometry}
\usepackage{float}
\usepackage{amsmath}
\usepackage{amssymb}
\usepackage{mathdots}

\makeatletter

\providecommand{\tabularnewline}{\\}

\usepackage[english]{babel}
\usepackage{txfonts}\usepackage[classicReIm]{kpfonts}%%% use 'pdftex' instead of 'dvips' for PDF output

\makeatother

\usepackage{babel}
\begin{document}
%\selectlanguage{english} %%% remove comment delimiter ('%') and select language if required

\title{\noindent A Study of the Time Evolution of Brans-Dicke Parameter
and its Role in Cosmic Expansion}

\author{\textbf{Sudipto Roy}$^{\mathbf{1}}$\textbf{, Soumyadip Chowdhury}$^{\mathbf{2}}$ }

\maketitle
\begin{center}
$^{\mathbf{1}}$Department of Physics, St. Xavier's College, Kolkata 
\par\end{center}

\begin{center}
$^{\mathbf{2}}$Student of M.Sc. (Physics), Batch of 2015-17, St.
Xavier's College, Kolkata 
\par\end{center}

\begin{center}
30 Mother Teresa Sarani (Park Street), Kolkata -- 700016, West Bengal,
India. 
\par\end{center}

\begin{center}
E-mail: $^{\mathbf{1}}$roy.sudipto1@gmail.com, $^{\mathbf{2}}$soumyadip.physics@gmail.com 
\par\end{center}
\begin{abstract}
\noindent The dependence of Brans-Dicke (BD) parameter upon the scalar
field, for different cosmological era of the expanding universe, has
been explored. The time dependence of the scalar field has been determined
and thereby the explicit time dependence of the BD parameter has been
obtained. Experimental observations regarding the time dependence
of the gravitational constant has been considered for this study.
The scale factor, Hubble parameter, deceleration parameter, matter
density, gravitational constant and the density parameters for matter
and dark energy have been expressed in terms of the BD parameter and
its time derivatives, showing its role in cosmic expansion.
\end{abstract}
\noindent Keywords: Cosmology, Brans-Dicke theory, Scalar Field, Accelerated
cosmic expansion, Gravitational Constant, Density parameters for matter
and dark energy.

\noindent PACS Numbers: 04.20.-q, 98.80.Jk, 98.80.-k, 95.30.Sf, 95.30.Ft,
91.10.Op, 95.36.+x

\section{\noindent Introduction}

\noindent The theory of gravitation, known as Brans-Dicke (BD) theory,
is characterized by a dimensionless constant function $\omega$ and
a scalar field $\varphi$. The arbitrary coupling function $\omega$
determines the relative importance {[}1{]}. In the generalized version
of the B-D theory, the dimensionless coupling constant is considered
to a be a function of time {[}2{]}. This time dependence may also
be expressed by assuming $\omega$ to be explicitly dependent upon
the scalar field $\varphi$ {[}3{]}. There are several important reasons
for which generalised BD theory has gained so much of important in
explaining and analyzing cosmological phenomena. In Kaluza-Klein theories,
super gravity theory and in all the known effective string actions,
this theory has a natural appearance. It is regarded as the most natural
extension of the General Theory of Relativity (GTR), which may justify
its applicability in fundamental theories {[}4{]}. In the generalized
version of Brans-Dicke theory, which is also known as graviton-dilaton
theory, $\omega$ has been shown to be a function of the scalar field
$\varphi$ (dilaton). Thus, there can be several models depending
upon the functional form of the BD parameter. This theory generates
the results obtained from GTR, for a constant scalar field and an
infinite $\omega$ {[}5, 6{]}. Using a constant value of $\omega$,
BD theory was found to account for almost all important cosmological
observations regarding the evolution of the universe. BD theory is
capable of explaining the features like inflation {[}7{]}, early and
late time behaviour of the universe {[}8{]}, cosmic acceleration and
structure formation {[}9{]}, quintessence and the coincidence problem
{[}10{]}, self -interacting potential and cosmic acceleration {[}11{]}.
For a small negative value of $\omega$ it correctly explains cosmic
acceleration, structure formation and the coincidence problem and,
for a large value of $\omega$, BD theory gives the correct amount
of inflation and early and late time behaviour. The time dependence
of $\omega$ in Brans-Dicke theory has many interesting features.
From string and Kaluza-Klein theories it gets a strong corroboration,
and in several studies, the dynamics of the universe has been analyzed
within its framework. Through these attempts, the phenomena like the
evolution of the universe, its accelerated expansion and quintessence,
have been explained in a qualitative way without deriving any explicit
time dependence of the BD parameter {[}10, 12-14{]}. Some recent studies
have shown that several models can be formulated based on the concept
of a time-varying $\omega$ {[}6{]}. Therefore it would be quite natural
for the researchers to find an analytical expression of $\omega$
as a function of time, using the field equations of the Brans-Dicke
theory. Some attempts were made to derive the time dependence of the
BD parameter using very simple empirical forms of scale factor and
the scalar field {[}15, 16{]}. But the chosen scale factors in these
cases produced time independent deceleration parameters, in complete
contradiction of observations.

\noindent The purpose of the present study is to derive an expression
of the BD parameter as an explicit function of the scalar field ($\varphi$)
and also an expression showing its explicit time dependence, for different
cosmological era of the universe, characterized by different values
of the equation of state parameter ($\gamma$). The present matter-dominated
universe contains cold matter of negligible pressure (dust), with
$\gamma=0$. For the present study, empirical expressions of the scale
factor, scalar field and the BD parameter have been used. Using the
field equations of BD theory and also the wave equation for the BD
scalar field ($\varphi$), we have determined the values of all parameters
involved in these empirical forms. We have also used the experimental
results regarding the time variation of the gravitational constant
($G$), to determine the values of these unknown parameters. The scale
factor ($a$), Hubble parameter ($H$), deceleration parameter ($q$)
matter density ($\rho$) and gravitational constant ($G$) have been
expressed as functions of the BD parameter ($\omega$) and its time
derivatives. The dependence of density parameters, for matter and
dark energy (${\mathrm{\Omega}}_{m}\ ,\ {\mathrm{\Omega}}_{d}$),
upon the BD parameter have also been shown theoretically. These expressions
show mathematically, the role played by the dimensionless parameter
($\omega$) in cosmic expansion.

\section{\noindent Theoretical Model}

\noindent In generalized Brans-Dicke theory, the field equations,
for a universe filled with a perfect fluid and described by Friedmann-Robertson-Walker
space-time, with scale factor $a(t)$ and spatial curvature $k$,
are expressed as,

\noindent 
\begin{align}
\frac{{\dot{a}}^{2}+k}{a^{2}}+\frac{\dot{a}}{a}\frac{\dot{\varphi}}{\varphi}-\frac{\omega\left(\varphi\right)}{6}\frac{{\dot{\varphi}}^{2}}{{\varphi}^{2}} & =\frac{\rho}{3\varphi}\label{GrindEQ__01_}
\end{align}

\begin{align}
2\frac{\ddot{a}}{a}+\frac{{\dot{a}}^{2}+k}{a^{2}}+\frac{\omega\left(\varphi\right)}{2}\frac{{\dot{\varphi}}^{2}}{{\varphi}^{2}}+2\frac{\dot{a}}{a}\frac{\dot{\varphi}}{\varphi}+\frac{\ddot{\varphi}}{\varphi} & =-\frac{P}{\varphi}\ \ \ \ \ \ \ \ \label{GrindEQ__02_}
\end{align}

\noindent The wave equation for the scalar field ($\varphi$), in
generalized Brans-Dicke theory, where $\omega$ is a time dependent
parameter, is expressed as,

\noindent 
\begin{align}
\ddot{\varphi}+3\frac{\dot{a}\dot{\varphi}}{a} & =\frac{\rho-3P}{2\omega+3}-\frac{\dot{\omega}\dot{\varphi}}{2\omega+3}\label{GrindEQ__03_}
\end{align}

\noindent The energy conservation for the cosmic fluid is given by
{[}15, 16{]},

\noindent 
\begin{align}
\dot{\rho}+3\frac{\dot{a}}{a}\left(\rho+P\right) & =0\label{GrindEQ__04_}
\end{align}

\noindent The equation of state of the fluid is expressed as,

\noindent 
\begin{align}
P & =\gamma\rho\label{GrindEQ__05_}
\end{align}

\noindent The values of $\gamma$ are $-1$ (vacuum energy dominated
era), $0$ (matter dominated era), ${1}/{3}$ (radiation dominated
era), $1$ (massless scalar field dominated era).

\noindent The solution of equation (4), using equation (5), is obtained
as,

\noindent 
\begin{align}
\rho & ={\rho}_{0}a^{-3(1+\gamma)}\label{GrindEQ__06_}
\end{align}

\noindent For the present study, based on the equations (1), (2) and
(3), the following empirical relations of scale factor, scalar field
and BD parameter have been chosen.

\noindent 
\begin{align}
a & =a_{0}{\ \ \left({t}/{t_{0}}\right)}^{\varepsilon}\ \ Exp\left[\mu\left(t-t_{0}\right)\right]\label{GrindEQ__07_}
\end{align}
\begin{align}
\varphi & ={\varphi}_{0}{\left({a}/{a_{0}}\right)}^{n}\label{GrindEQ__08_}
\end{align}
\begin{align}
\omega & ={\omega}_{0}Exp\ \left[m\left(\varphi-{\varphi}_{0}\right)\right]\label{GrindEQ__09_}
\end{align}

\noindent The scale factor (in eq. 7) has been chosen to ensure a
change of sign of the deceleration parameter with time, as per many
recent studies showing a transition of cosmic expansion from deceleration
to acceleration {[}3, 19, 20{]}. Here $\varepsilon,\mu>0$ to ensure
increase of scale factor with time. The Hubble parameter ($H$) and
the deceleration parameter ($q$), calculated from this scale factor,
are written below.

\noindent 
\begin{align}
H & =\mu+\frac{\varepsilon}{t}\label{GrindEQ__10_}
\end{align}
\begin{align}
q & =-1+\frac{\varepsilon}{{\left(\varepsilon+\mu t\right)}^{2}}\ \label{GrindEQ__11_}
\end{align}

\noindent For $0<\varepsilon<1$, we get $q>0$ at $t=0$ and, for
$t?8$, we have $q?-1$.

\noindent Taking $H=H_{0}$ and $q=q_{0}$, at $t=t_{0}$, one gets,

\noindent 
\begin{align}
\varepsilon & ={\left(H_{0}t_{0}\right)}^{2}\ \left(q_{0}+1\right)\label{GrindEQ__12_}
\end{align}
\begin{align}
\mu & =H_{0}-{H_{0}}^{2}t_{0}\left(q_{0}+1\right)\label{GrindEQ__13_}
\end{align}

\noindent The scalar field (in eq. 8) has been chosen on the basis
of some studies on Brans-Dicke theory {[}3, 19, 20{]}. The value of
$n$ can be determined from the field equations.

\noindent The empirical expression of BD parameter (in eq. 9) has
been chosen according to the generalized Brans-Dicke theory, where
$\omega$ is regarded as a function of the scalar field ($\varphi$)
{[}3{]}. The values of ${\omega}_{0}$ and $m$ have to be determined
from the field equations.

\noindent Considering $\omega$ as a function of $\varphi$, equation
(3) can be written as,

\noindent 
\begin{align}
\ddot{\varphi}+3\frac{\dot{a}\dot{\varphi}}{a} & =\frac{\rho-3P}{2\omega+3}-\frac{{\dot{\varphi}}^{2}}{2\omega+3}\frac{d\omega}{d\varphi}\label{GrindEQ__14_}
\end{align}

\noindent Combining the equations (5), (8), (9) with (14) one gets,

\noindent 
\begin{align}
\omega & =\frac{1}{f\ H^{2}}\ \left[\frac{\rho\left(1-3\gamma\right)}{\varphi}-3H^{2}\ \left(n^{2}+2n-nq\right)\right]\label{GrindEQ__15_}
\end{align}

\noindent Where, 
\begin{align*}
f\left(t\right) & =2\left(n^{2}+2n-nq\right)+mn^{2}\varphi
\end{align*}
 
\[
H(t)=\mu+\frac{\varepsilon}{t}
\]
\[
q(t)=-1+\frac{\varepsilon}{{\left(\varepsilon+\mu t\right)}^{2}}
\]
\[
\rho(t)=\frac{{\rho}_{0\ }}{{a_{0}}^{3\left(1+\gamma\right)\ }}\ {\ \left(\frac{t}{t_{0}}\right)}^{-3\varepsilon\left(1+\gamma\right)}\ \ Exp\left[-3\left(1+\gamma\right)\mu\left(t-t_{0}\right)\right]
\]
\[
\varphi(t)={\varphi}_{0\ \ }{\ \left(\frac{t}{t_{0}}\right)}^{n\varepsilon}\ \ Exp\left[n\mu\left(t-t_{0}\right)\right]
\]

\noindent The two above equations, for $\rho$ and $\varphi$, are
obtained by substituting equation (7) into the equations (6) and (8)
respectively.

\noindent Taking $\gamma=0$ in equation \eqref{GrindEQ__15_} for
the present matter dominated era and writing all parameter values
for the present time $\left(t=t_{0}\right)$, one obtains,

\noindent 
\begin{align}
m & =\frac{{\rho}_{0}}{n^{2}{\omega}_{0}{H_{0}}^{2}\ {{\varphi}_{0}}^{2}}-\frac{\left(2{\omega}_{0}+3\right)\left(n^{2}+2n-nq_{0}\right)}{n^{2}{\omega}_{0}{\varphi}_{0}}\label{GrindEQ__16_}
\end{align}

\noindent Using the equations (1), (2) and (8), and taking $k=0$,
$P=0$, we get,

\noindent 
\begin{align}
\omega & =\frac{1}{n^{2}}\left[2+2q+2n-n^{2}+nq-\frac{\rho}{\varphi H^{2}}\right]\label{GrindEQ__17_}
\end{align}

\noindent Taking all parameter values at $t=t_{0}$ in equation \eqref{GrindEQ__17_},
one gets,

\noindent 
\begin{align}
{\omega}_{0} & =\frac{1}{n^{2}}\left[2+2q_{0}+2n-n^{2}+nq_{0}-\frac{{\rho}_{0}}{{\varphi}_{0\ }{H_{0}}^{2}}\right]\label{GrindEQ__18_}
\end{align}

\noindent Using equation \eqref{GrindEQ__18_} in \eqref{GrindEQ__16_}
one obtains,

\noindent 
\[
m=\frac{{\rho}_{0\ }}{\left[2\left(1+q_{0}+n\right)-n^{2}+nq_{0}-\frac{{\rho}_{0}}{{\varphi}_{0\ }{H_{0}}^{2}}\right]{H_{0}}^{2}\ {{\varphi}_{0}}^{2}}
\]
\begin{align}
\ \ \ \ \ \ \ \ \ \ -\ \frac{\left[\frac{2}{n^{2}}\left\{ 2\left(1+q_{0}+n\right)-n^{2}+nq_{0}-\frac{{\rho}_{0}}{{\varphi}_{0\ }{H_{0}}^{2}}\right\} +3\right]\left(n^{2}+2n-nq_{0}\right)}{\left[2\left(1+q_{0}+n\right)-n^{2}+nq_{0}-\frac{{\rho}_{0}}{{\varphi}_{0\ }{H_{0}}^{2}}\right]\ {\varphi}_{0}}\label{GrindEQ__19_}
\end{align}

\noindent This expression of $m$ (eqn. 19) should be substituted
into the expression of $f(t)$ in the expression of $\omega$ in equation
\eqref{GrindEQ__15_}.

\noindent Eliminating $\omega$ from the equations (1) and (2), taking
$k=0$ and $P=\gamma\rho$ one obtains,

\noindent 
\begin{align}
2\frac{\ddot{a}}{a}+4\frac{{\dot{a}}^{2}}{a^{2}}+5\frac{\dot{a}}{a}\frac{\dot{\varphi}}{\varphi}+\frac{\ddot{\varphi}}{\varphi} & =\frac{\rho}{\varphi}\left(1-\gamma\right)\label{GrindEQ__20_}
\end{align}

\noindent Using equation (8) in (20) for $\gamma=0$ and taking all
parameter values at $t=t_{0}$ one gets,

\noindent 
\begin{align}
n^{2}+\left(4-q_{0}\right)n+\left(4-2q_{0}-\frac{{\rho}_{0}}{{\varphi}_{0}{H_{0}}^{2}}\right) & =0\label{GrindEQ__21_}
\end{align}

\noindent Equation \eqref{GrindEQ__21_} is quadratic in $n$. Its
two roots are given by,

\noindent 
\begin{align}
n_{\pm} & =\frac{1}{2}\left[q_{0}-4\pm{\left({q_{0}}^{2}+\frac{4{\rho}_{0}}{{\varphi}_{0}{H_{0}}^{2}}\right)}^{1/2}\right]\label{GrindEQ__22_}
\end{align}

\noindent The values of different cosmological parameters used in
this article are:

\noindent 
\[
H_{0}=\frac{72\frac{Km}{s}}{Mpc}=2.33\times{10}^{-18}\ {sec}^{-1},q_{0}=-0.55,{\rho}_{0}=2.83\times{10}^{-27}Kg\ m^{-3}
\]
\[
{\varphi}_{0}=\frac{1}{G_{0}}=1.498\times{10}^{10}{Kg}^{2}\ m^{-2}N^{-1},t_{0}=\ 4.36\times{10}^{17}\ s
\]

\noindent Using these values in equation \eqref{GrindEQ__22_} one
gets,

\noindent 
\begin{align}
n_{\pm} & =-1.94,-2.61
\end{align}
Since $\varphi={\varphi}_{0}a^{n}$ was chosen empirically (and not
obtained as a solution of the field equations) the parameter $n$
can also take values other than the two values shown by equation (23).
The important fact about these two values is that both of them are
negative, indicating a decrease of $\varphi$ and an increase of $G(=\frac{1}{\varphi})$
with time. There are studies that show a decrease of $\varphi$ with
time {[}3, 15{]}.

\noindent According to a study by Banerjee and Pavon {[}10{]}, $-3/2<{\omega}_{0}<0$.

\noindent Using equation \eqref{GrindEQ__18_}, the ranges of $n$
values, satisfying this requirement, are found to be,

\noindent 
\begin{align}
n & <-2.061,-0.838<n<-0.455,n>1.905
\end{align}
The sign of $n$ determines whether $\varphi$ increases or decreases
with time, as per equation (8) where the scale factor $a(t)$ always
increases with time in an expanding universe. The larger the value
of $\left|n\right|$, greater would be its rate of change with time.
Using equation (8), the fractional rate of change of the gravitational
constant is given by,

\noindent 
\begin{align*}
{\left(\frac{\dot{G}}{G}\right)}_{t=t_{0}} & ={\left[\frac{1}{1/\varphi}\frac{d}{dt}\left(\frac{1}{\varphi}\right)\right]}_{t=t_{0}}=-{\left(\frac{\dot{\varphi}}{\varphi}\right)}_{t=t_{0}}=-nH_{0}
\end{align*}

\noindent or, 
\begin{align}
n & =-\frac{1}{H_{0}}{\left(\frac{\dot{G}}{G}\right)}_{t=t_{0}}
\end{align}
Using equation (25), the values of $n$ can be more reliably determined
from the experimental findings of ${\left(\frac{\dot{G}}{G}\right)}_{t=t_{0}}$.
Its sign is found to be both positive and negative experimentally
{[}17{]}.

\noindent Combining equation (8) with (20), one obtains,

\noindent 
\begin{align}
\varphi & =\frac{\rho(1-\gamma)}{H^{2}\left(2+n\right)(2+n-q)}
\end{align}
 Using equation (26), one obtains the expression for ${\varphi}_{0}$
by taking all parameter values at $t=t_{0}$ and $\gamma=0$ for the
present matter dominated era of the universe.

\noindent 
\begin{align}
{\varphi}_{0} & =\frac{{\rho}_{0}}{{H_{0}}^{2}\left(2+n\right)(2+n-q_{0})}
\end{align}

\noindent To express the dependence of $\omega$ upon $\gamma$, equations
(26, 27) and \eqref{GrindEQ__18_} are substituted into equation (9),
leading to the following equation.

\noindent 
\[
\omega=\frac{1}{n^{2}}\left[2+2q_{0}+2n-n^{2}+nq_{0}-\frac{{\rho}_{0}}{{\varphi}_{0\ }{H_{0}}^{2}}\right]
\]
\begin{align}
\ \ \ \ \ \ \ \ \ \times Exp\ \left[m\left\{ \frac{\rho(1-\gamma)}{H^{2}\left(2+n\right)(2+n-q)}-\frac{{\rho}_{0}}{{H_{0}}^{2}\left(2+n\right)(2+n-q_{0})}\right\} \right]\label{GrindEQ__27_}
\end{align}

\noindent To express the explicit time dependence of $\omega$, substitutions
have to be made from the equations (6), (7), (10), (11) into equation
(28), leading to the following form.

\noindent 
\begin{align}
\omega & =A\ Exp\left[\frac{m{\rho}_{0}{\left[a_{0}{\left(\frac{t}{t_{0}}\right)}^{\varepsilon}Exp\left\{ \mu\left(t-t_{0}\right)\right\} \right]}^{-3\left(1+\gamma\right)}\left(1-\gamma\right)}{{\left(\mu+\frac{\varepsilon}{t}\right)}^{2}\left(2+n\right)\left[2+n-\left\{ -1+\frac{\varepsilon}{{\left(\varepsilon+\mu t\right)}^{2}}\right\} \right]}\right]\label{GrindEQ__28_}
\end{align}

\noindent where $A={\omega}_{0}\ Exp\left[\frac{-m{\rho}_{0}}{{H_{0}}^{2}\left(2+n\right)(2+n-q_{0})}\right]$

\noindent Here, $\varepsilon$, $\mu$, ${\omega}_{0}$, $m$ and
$n$ are given by equations \eqref{GrindEQ__12_}, \eqref{GrindEQ__13_},
\eqref{GrindEQ__18_}, \eqref{GrindEQ__19_} and (25) respectively.

\noindent Using the equations (6-9), one obtains the following expressions
of different cosmological quantities of interest in terms of $\omega$
and its derivatives.

\noindent 
\begin{align}
a & =a_{0}{\left[1+\frac{1}{m{\varphi}_{0}}\ ln\left(\frac{\omega}{{\omega}_{0}}\right)\right]}^{1/n}\label{GrindEQ__29_}
\end{align}

\begin{align}
H & =\frac{\dot{a}}{a}=\frac{1}{mn{\varphi}_{0}}\ {\left[1+\frac{1}{m{\varphi}_{0}}\ ln\left(\frac{\omega}{{\omega}_{0}}\right)\right]}^{-1}\frac{\dot{\omega}}{\omega}\label{GrindEQ__30_}
\end{align}

\begin{align}
q & =-\frac{\ddot{a}a}{{\dot{a}}^{2}}=-1+n\left[1+\left\{ m{\varphi}_{0}+ln\left(\frac{\omega}{{\omega}_{0}}\right)\right\} \left(1-\frac{\ddot{\omega}\omega}{{\dot{\omega}}^{2}}\right)\right]\label{GrindEQ__31_}
\end{align}

\begin{align}
\rho & ={\rho}_{0\ }{a_{0}}^{-3\left(1+\gamma\right)}{\ \left[1+\frac{1}{m{\varphi}_{0}}\ ln\left(\frac{\omega}{{\omega}_{0}}\right)\right]}^{\frac{-3\left(1+\gamma\right)}{n}}\label{GrindEQ__32_}
\end{align}

\begin{align}
G & =\frac{1}{\varphi}=\frac{1}{{\varphi}_{0}+\left(1/m\right)\ ln\left(\omega/{\omega}_{0}\right)}\label{GrindEQ__33_}
\end{align}

\noindent In the above expressions (30-34), the parameters ${\omega}_{0}$
and $m$ (eqns. 18, 19) are functions of the parameter $n$ which
controls the change of the scalar field ($\varphi$) with time.

\noindent Using equation (33), the density parameter for all matter
(dark + baryonic) can be written as,

\noindent 
\begin{align}
\Omega_{m} & =\frac{\rho}{{\rho}_{c}}=\frac{{\rho}_{0}}{{\rho}_{c}}{a_{0}}^{-3\left(1+\gamma\right)}{\ \left[1+\frac{1}{m{\varphi}_{0}}\ ln\left(\frac{\omega}{{\omega}_{0}}\right)\right]}^{\frac{-3\left(1+\gamma\right)}{n}}\label{GrindEQ__34_}
\end{align}

\noindent Here, ${\rho}_{c}$ is the critical density of matter-energy
of the universe. ${\rho}_{c}\cong{10}^{-26}Kg\ m^{-3}$

\noindent The density parameter for dark energy is given by,

\noindent 
\begin{align}
\Omega_{d} & =1-O_{m}=1-\frac{{\rho}_{0}}{{\rho}_{c}}{a_{0}}^{-3\left(1+\gamma\right)}{\ \left[1+\frac{1}{m{\varphi}_{0}}\ ln\left(\frac{\omega}{{\omega}_{0}}\right)\right]}^{\frac{-3\left(1+\gamma\right)}{n}}\label{GrindEQ__35_}
\end{align}

\noindent Equations (35) and (36) show the dependence of density parameters,
of matter and dark energy respectively, upon the Brans-Dicke parameter.

\noindent The present era of the universe is matter dominated with
negligible pressure ($\gamma=0$). From the equations (6) and (7),
we thus have, 
\begin{align}
\rho(t) & =\frac{{\rho}_{0\ }}{{a_{0}}^{3\ }}\ {\ \left(\frac{t}{t_{0}}\right)}^{-3\varepsilon}\ \ Exp\left[-3\mu\left(t-t_{0}\right)\right]\label{GrindEQ__36_}
\end{align}

\noindent For the present universe, the expressions of $\omega$,
from the equations (15) and (29), are respectively obtained as the
following two equations, (38) and (39).

\noindent 
\begin{align}
\omega & =\frac{1}{f\ H^{2}}\ \left[\frac{\rho}{\varphi}-3H^{2}\ \left(n^{2}+2n-nq\right)\right]\label{GrindEQ__37_}
\end{align}

\begin{align}
\omega & =A\ Exp\left[\frac{m{\rho}_{0}{\left[a_{0}{\left(\frac{t}{t_{0}}\right)}^{\varepsilon}Exp\left\{ \mu\left(t-t_{0}\right)\right\} \right]}^{-3}}{{\left(\mu+\frac{\varepsilon}{t}\right)}^{2}\left(2+n\right)\left[2+n-\left\{ -1+\frac{\varepsilon}{{\left(\varepsilon+\mu t\right)}^{2}}\right\} \right]}\right]\label{GrindEQ__38_}
\end{align}

\noindent with $A={\omega}_{0}\ Exp\left[\frac{-m{\rho}_{0}}{{H_{0}}^{2}\left(2+n\right)(2+n-q_{0})}\right]$

\noindent In equation (38), $f\left(t\right)=2\left(n^{2}+2n-nq\right)+mn^{2}\varphi$
and $\rho(t)$ is given by equation (37).

\noindent Combining the equations (8) and (9) one gets,

\noindent 
\begin{align}
\omega & ={\omega}_{0}Exp\ \left[m{\varphi}_{0}\left\{ {\left({a}/{a_{0}}\right)}^{n}-1\right\} \right]\label{GrindEQ__39_}
\end{align}

\noindent Using equations (7) and \eqref{GrindEQ__18_} in (40) one
obtains,

\noindent 
\begin{align}
\omega & =\frac{1}{n^{2}}\left[2+2q_{0}+2n-n^{2}+nq_{0}-\frac{{\rho}_{0}}{{\varphi}_{0\ }{H_{0}}^{2}}\right]\times Exp\ \left[m{\varphi}_{0}\left\{ {\left({t}/{t_{0}}\right)}^{n\varepsilon}\ \ Exp\left[n\mu\left(t-t_{0}\right)\right]-1\right\} \right]
\end{align}

\noindent Equation (41) is an expression of $\omega$ without any
explicit dependence on $\gamma$ and thus applicable for the entire
span of cosmological time of the expanding universe. Equations \eqref{GrindEQ__12_},
\eqref{GrindEQ__13_}, \eqref{GrindEQ__19_} and (25) can be used
for $\varepsilon,\ \mu,m,n$ respectively.

\noindent 

\noindent

\section{\noindent Results}

\noindent 

\noindent Table-1 has a list of values of $n$, ${\omega}_{0}$ and
$m$ for different values of ${\left(\frac{\dot{G}}{G}\right)}_{t=t_{0}}$
whose range has been chosen from experimental findings {[}17-20{]}.
Both ${\omega}_{0}$ and $m$ depend upon $n\ $(eqns 18, 19). The
signs of the parameters, $n$ and $m$, determine whether $\omega$
increases or decreases with time, as per equations (8) and (9). When
they have the same sign, $\omega$ increases with time if ${\omega}_{0}$
is positive. When they have opposite signs, $\omega$ decreases with
time if ${\omega}_{0}$ is positive. For these two cases, the reverse
happens if ${\omega}_{0}$ is negative. Table-1 shows that for extremely
low values of $\left|{\left(\frac{\dot{G}}{G}\right)}_{t=t_{0}}\right|$,
one gets very large values of $\left|{\omega}_{0}\right|$. It clearly
indicates that larger values of $\omega$ causes smaller rate of change
of the gravitational constant $\left(G=1/\varphi\right)$. Equation
\eqref{GrindEQ__15_} shows a linear relation between $\frac{d\omega}{dt}$
and $\frac{d\rho}{dt}$ and equation (6) suggests that $\rho$ decreases
faster for larger values of $\gamma$. Thus, one concludes that $\omega$
changes faster for greater values of $\gamma$.

\section{\noindent Conclusion}

\noindent 

\noindent In the present study we have determined the nature of dependence
of the Brans-Dicke parameter upon the scalar field and also the dependence
of the scalar field upon time. Therefore, this study has enabled us
to determine the time dependence of BD parameter which plays a very
important role in the accelerated expansion of the universe. The signs
of the parameters $n$, $m$ and ${\omega}_{0}$ determine whether
the BD parameter increases or decreases with time. Actually, both
${\omega}_{0}$ and $m$ are functions of $n$ (eqns 18, 19), which
determines the time dependence of the gravitational constant (eqn.
25). According to a study by Banerjee and Pavon {[}10{]}, we must
have $-3/2<{\omega}_{0}<0$. The range of values for $n$, satisfying
this requirement, have been determined. The values of $n$ can be
best determined by the experimentally measured values of ${\left(\frac{\dot{G}}{G}\right)}_{t=t_{0}}$,
reported by several researchers {[}17-20{]}. For a wide range of values
of this quantity, we have determined the values of the parameters
$n$, $m$ and ${\omega}_{0}$ and listed them in Table-1. In deriving
the expression of ${\omega}_{0}$ (eqn. 18), equal weightages have
been given to the equations (1) and (2). Instead of doing this, one
may calculate the value of ${\omega}_{0}$ separately from the equations
(1) and (2) and determine their weighted average with unequal weights
assigned to these two values. For this purpose, a new parameter can
be introduced to represent their relative weightage, allowing us to
determine the value of ${\omega}_{0}$ correctly, making it more consistent
with more advanced studies in this regard. Using the same set of empirical
relations (eqns. 8 and 9) for the scalar field and the BD parameter,
we have determined three expressions of $\omega(t)$, represented
by the equations (15), (29) and (41), showing clearly its dependence
upon time and the equation of state parameter ($\gamma$). The form
of these expressions for the present matter dominated state of the
universe, which is considered to be a pressureless dust with $\gamma=0$,
are shown by the equations (38) and (39). A limitation of this study
is that, it is based upon empirical forms of the scale factor $\left(a\right)$
and the scalar field $\left(\varphi\right)$, which are not the solutions
of the Brans-Dicke field equations. An improvement over this method
can be made by choosing one of these parameters empirically and determining
the other from the field equations. This is our plan for a completely
new study to be carried out in future.

\noindent 

\noindent

\section{\noindent Acknowledgement}

\noindent The authors are thankful to many academicians and researchers
whose contributions to this field have enriched them and also encouraged
them intellectually.

\begin{table}[H]

\begin{centering}
\textbf{Table-1: Values of $n,\ {\omega}_{0},\ m$ for different values
of ${\left({\dot{G}}/{G}\right)}_{t=t_{0}}$}
\par\end{centering}

\begin{centering}
Here, 'E-10' means $\times10^{-10}$
\par\end{centering}

\centering{}%
\begin{tabular}{|c|c|c|c|}
\hline 
$\left(\frac{\dot{G}}{G}\right)_{t=t_{0}}Yr^{-1}$ & n & $\omega_{0}$ & m\tabularnewline
\hline 
\hline 
-1.0E-9 & 13.60935278  & -8.88784E-01  & 1.09005E-10 \tabularnewline
\hline 
-7.5E-10 & 10.20701459  & -8.49636E-01  & 1.27703E-10 \tabularnewline
\hline 
-5E-10 & 6.804676391  & -7.68226E-01  & 1.74769E-10 \tabularnewline
\hline 
-2.5E-10 & 3.402338196  & -4.99082E-01  & 4.68038E-10 \tabularnewline
\hline 
-1E-10 & 1.360935278  & 5.32569E-01  & -1.46195E-09 \tabularnewline
\hline 
-7.5E-11 & 1.020701459  & 1.25104E+00  & -1.02528E-09 \tabularnewline
\hline 
-5E-11 & 0.680467639  & 2.99939E+00  & -9.49143E-10 \tabularnewline
\hline 
-2.5E-11 & 0.34023382  & 1.07358E+01  & -1.29075E-09 \tabularnewline
\hline 
-1E-11 & 0.136093528  & 5.63670E+01  & -2.70303E-09 \tabularnewline
\hline 
-7.5E-12 & 0.102070146  & 9.62505E+01  & -3.52075E-09 \tabularnewline
\hline 
-5E-12 & 0.068046764  & 2.07159E+02  & -5.17152E-09 \tabularnewline
\hline 
-2.5E-12 & 0.034023382  & 7.89019E+02  & -1.01567E-08 \tabularnewline
\hline 
-1E-12 & 0.013609353  & 4.77680E+03  & -2.51550E-08 \tabularnewline
\hline 
-7.5E-12 & 0.010207015  & 8.44551E+03  & -3.34917E-08 \tabularnewline
\hline 
-5E-13 & 0.006804676  & 1.88971E+04  & -5.01672E-08 \tabularnewline
\hline 
-2.5E-13 & 0.003402338  & 7.51653E+04  & \raggedright{}-1.00198E-07 \tabularnewline
\hline 
-1E-13 & 0.001360935  & 4.68190E+05  & -2.50293E-07 \tabularnewline
\hline 
1E-13 & -0.001360935  & 4.66059E+05  & 2.50032E-07 \tabularnewline
\hline 
2.5E-13 & -0.003402338  & 7.43129E+04  & 9.99359E-08 \tabularnewline
\hline 
5E-13 & -0.006804676  & 1.84709E+04  & 4.99056E-08 \tabularnewline
\hline 
7.5E-13 & -0.010207015  & 8.16139E+03  & 3.32302E-08 \tabularnewline
\hline 
1E-12 & -0.013609353  & 4.56371E+03  & 2.48936E-08 \tabularnewline
\hline 
2.5E-12 & -0.034023382  & 7.03783E+02  & 9.89685E-09 \tabularnewline
\hline 
5E-12 & -0.068046764  & 1.64541E+02  & 4.91717E-09 \tabularnewline
\hline 
7.5E-12 & -0.102070146  & 6.78386E+01  & 3.27607E-09 \tabularnewline
\hline 
1E-11 & -0.136093528  & 3.50581E+01  & 2.47301E-09 \tabularnewline
\hline 
2.5E-11 & -0.34023382  & 2.21223E+00  & 1.46417E-09 \tabularnewline
\hline 
5E-11 & -0.680467639  & -1.26239E+00  & -7.30199E-11 \tabularnewline
\hline 
7.5E-11 & -1.020701459  & -1.59015E+00  & 9.93734E-12 \tabularnewline
\hline 
1E-10 & -1.360935278  & -1.59832E+00  & 6.39048E-12 \tabularnewline
\hline 
2.5E-10 & -3.402338196  & -1.35144E+00  & 3.52820E-12 \tabularnewline
\hline 
5E-10 & -6.804676391  & -1.19440E+00  & 2.13166E-11 \tabularnewline
\hline 
7.5E-10 & -10.20701459  & -1.13375E+00  & 3.23346E-11 \tabularnewline
\hline 
1E-9 & -13.60935278  & -1.10187E+00  & 3.91899E-11 \tabularnewline
\hline 
\end{tabular}
\end{table}

\end{document}